\documentclass[preprint,12pt]{elsarticle}

\usepackage{amssymb}
\usepackage{amsmath}
\usepackage{soul}
\usepackage{xcolor}

\begin{document}

\begin{frontmatter}



\title{Augmentation-Based Deep Learning for Identification of Circulating Tumor Cells }
%{\footnotesize \textsuperscript{*}Note: Sub-titles are not captured for https://ieeexplore.ieee.org  and

%% Author name
\author[label1]{Martina Russo} 
\author[label2]{Giulia Bertolini} 
\author[label2]{Vera Cappelletti} 
\author[label2]{Cinzia De Marco} 
\author[label2]{Serena Di Cosimo} 
\author[label3]{Petra Paiè} 
\author[label1]{Nadia Brancati} 
\affiliation[label1]{organization={Institute for High Performance Computing and Networking-National Research Council of Italy (ICAR-CNR)},%Department and Organization
            %addressline={}, 
            city={Naples},
            %postcode={}, 
            state={Italy},
            %country={}
            }

\affiliation[label2]{organization={Fondazione IRCCS-Istituto Nazionale dei Tumori},%Department and Organization
            %addressline={}, 
            city={Milan},
            %postcode={}, 
            state={Italy},
            %country={}
            }

\affiliation[label3]{organization={Politecnico di Milano},%Department and Organization
            %addressline={}, 
            city={Milan},
            %postcode={}, 
            state={Italy},
            %country={}
            }
%Traditional approaches rely on immunostaining techniques to detect specific markers, but these methods can be time-consuming, require expert interpretation, and may not capture the full diversity of CTC phenotypes \cite{zhang2022epithelial}.To overcome these limitations, unbiased enrichment and isolation techniques have been developed.

%% Abstract
\begin{abstract}

Circulating tumor cells (CTCs) are crucial biomarkers in liquid biopsy, offering a noninvasive tool for cancer patient management. However, their identification remains particularly challenging due to their limited number and heterogeneity. Labeling samples for contrast limits the generalization of fluorescence-based methods across different hospital datasets. Analyzing single-cell images enables detailed assessment of cell morphology, subcellular structures, and phenotypic variations, often hidden in clustered images. Developing a method based on bright-field single-cell analysis could overcome these limitations. CTCs can be isolated using an unbiased workflow combining Parsortix® technology, which selects cells based on size and deformability, with DEPArray™ technology, enabling precise visualization and selection of single cells. Traditionally, DEPArray-acquired digital images are manually analyzed, making the process time-consuming and prone to variability. In this study, we present a Deep Learning-based (DL) classification pipeline designed to distinguish CTCs from leukocytes in blood samples, aimed to enhance diagnostic accuracy and optimize clinical workflows. Our approach employs images from the bright-field channel acquired through DEPArray technology leveraging a ResNet-based Convolutional Neural Network. To improve model generalization, we applied three types of data augmentation techniques and incorporated fluorescence (DAPI) channel images into the training phase, allowing the network to learn additional CTC-specific features. Notably, only bright-field images have been used for testing, ensuring the model's ability to identify CTCs without relying on fluorescence markers. The proposed model achieved an F1-score of $0.798$, demonstrating its capability to distinguish CTCs from leukocytes. These findings highlight the potential of DL in refining CTC analysis and advancing liquid biopsy applications.

\end{abstract}


\end{frontmatter}
\section{\textbf{INTRODUCTION}}
Detection of cancer at an early stage significantly improves patient response to treatment and survival. However, nearly $50\%$ of patients are still diagnosed at an advanced stage~\cite{crosby2022early}. 
Recent advances in biological understanding and technological progress have led to the development of innovative approaches to improve early diagnosis. Single-cells sequencing and spatial transcriptomics shed new lights into the comprehension of tumor development and progression and in the interaction with tumor microenvironment. These data have improved precision therapy for cancer patients. However, the cost and time commitment are substantial, and the procedure requires a tumor tissue biopsy that can be invasive and, in some cases, not feasible for certain patients~\cite{han2022single}.
In this context, the analysis and detection of Circulating Tumor Cells (CTCs) is crucial. CTCs are cells shed from the primary tumor or metastatic sites into the bloodstream that have emerged as a valuable biomarker in oncology via liquid biopsy, providing a real-time and non-invasive means of cancer detection, monitoring, and treatment~\cite{alix2021liquid}.
Their extreme rarity and heterogeneity pose significant challenges for accurate identification and classification. Usually, techniques based on specific detection of markers expressed by CTC are performed, but they may fail to capture the full spectrum of the heterogeneous CTC population, resulting in an underestimation of CTC counts in other clinical settings~\cite{wang2020label}. The absence of a universally accepted method prevents the standardization and clinical validation of CTC-based biomarkers \cite{ju2022detection}. 
%In general, CTCs analysis is a challenging, intricate, and time-intensive manual task that requires experts to assess the images obtained from liquid biopsies individually.
%Therefore, the identification of CTCs remains a complex, time-consuming, and operator-dependent process, requiring expert manual evaluation. 
%Image analysis techniques could be used to develop methods for the automatic examination of CTC. 
%The ACCEPT Software is an image analysis package that through Computer Vision (CV) techniques automatize CTC classification, enumeration and phenotyping that is currently under active development for the EU Cancer-ID project~\cite{zeune2017quantifying}. Through these approaches it has been demonstrated that CTC size varies significantly across different tumor types and is generally smaller than tumor cell lines commonly used as references~\cite{mendelaar2021defining}. This underscores the importance of morphological analysis in optimizing size-based isolation methods. However, the need arises to isolate and acquire images of single cells so that their morphological characteristics can be studied. 

%In this context, CellSearch is fundamental since isolates CTCs by targeting the epithelial marker EpCAM and remains the only FDA-approved (Food and Drug Administration) system for enumerating CTCs in breast, colon, and prostate cancer clinical settings~\cite{riethdorf2018clinical}.
Consequently, there is a need for alternatives that can make this process more efficient and robust, facilitating the high-throughput application of CTC-based technologies and enhancing the study of cancer patients and the progression of their disease. CellSearch is the only FDA-approved (Food and Drug Administration) system for enumerating CTCs in different kinds of cancer clinical settings as breast, colon, and prostate cancer. This technology is based on capture of CTC expressing the epithelial marker EpCAM and therefore loses information on CTC that are not characterized by an epithelial phenotype~\cite{riethdorf2018clinical}. After the capture of CTC, a method to analyze images and classify them is required. The ACCEPT Software is an image analysis package for the automated CTC enumeration and phenotyping that is currently under active development for the EU Cancer-ID project~\cite{zeune2017quantifying}. The use of approaches of this kind has demonstrated that CTC size varies significantly across different tumor types and is generally smaller than tumor cell lines commonly used as references~\cite{mendelaar2021defining}. This underscores the importance of morphological analysis in optimizing size-based isolation methods and the need to acquire single-cell images of CTC to finely study their morphological characteristics. 
%This is fundamental to identifying a discriminating morphological component between CTCs and leukocytes. 
The ACCEPT Software is often used in combination with CellSearch for automatic analysis of CTCs. However, it relies on predefined rules and morphological criteria, which can introduce human bias and limit its adaptability to diverse imaging conditions. 

An alternative approach is a marker-independent strategies such as Parsortix technology~\cite{wishart2024molecular} that exploit a microfluid cassette that can enrich CTC based on their physical properties such as size and deformability. A notable innovation in single cell characterization and isolation of CTC is represented by the cell sorting technology based on imaging developed by the Di Trapani group, known as DEPArray™ (Menarini Silicon Biosystems, S.p.A., Italy) \cite{di2018deparray}. 
%The combination of Parsortix® technology, which isolates cells based on size and deformability, with DEPArray™ technology, which enables precise visualization and selection of single CTCs. 

Even with label-free enrichment technologies such as Parsortix®, 
%ClearCell®FX1, or filtration-based methods like ScreenCell®, 
CTCs may still be overlooked or misclassified as leukocytes when relying on immunofluorescence based evaluation~\cite{bates2023circulating}. This highlights the need for an alternative approach capable of identifying these cells based on their intrinsic morphological characteristics in bright-field (BF) imaging, reducing dependence on fluorescence markers. 
Despite these technologies' advantages, manual classification of CTCs remains a bottleneck in clinical workflows. This process requires expert analysis of digital images acquired through DEPArray, which is prone to inter-observer variability and may introduce subjectivity into the classification. 
In this context, Machine (ML) and Deep Learning (DL) models can learn directly from data, enabling more robust and automated CTC detection. They may excel at distinguishing CTCs from other blood cells, even in complex or suboptimal images, and can generalize across different datasets without extensive manual adjustments. %Additionally, deep networks offer higher accuracy, reduced error rates, and faster, scalable processing, making them more suitable for high-throughput applications.
In particular, Convolutional Neural Networks (CNNs), have demonstrated strong performance in automating CTC detection~\cite{guo2022circulating,vidlarova2023recent}, achieving high recall rates and reinforcing the potential of Artificial Intelligence (AI)-driven methods to enhance early cancer detection and monitoring in medical image analysis~\cite{painuli2022recent}.

In this study, we propose a DL-based classification pipeline designed to distinguish CTCs from leukocytes in a liquid biopsy using BF images acquired with DEPArray technology. In detail, a pre-trained CNN has been trained on a private dataset and given the small number of acquired CTC images, augmentation data have been used in the training phase. Augmentation operation include both affine and color transformation of BF images and also images coming from the DAPI fluorescence channel, which allows for improving performance of the model. Importantly, only BF images are used during testing, ensuring that the model can identify CTCs without reliance on fluorescence markers. 
%The aim is to contribute to the advancement of AI-driven methodologies for liquid biopsy analysis.

%The rest of the paper is organized as follows: previous related works are presented in Section \ref{rel}; a description of the proposed method and of the used data is presented in Section \ref{met}; the experimental setup, results and ablation studies are discussed in Section \ref{res}; and finally, in Section \ref{con} certain conclusions are drawn.

\section{\textbf{RELATED WORKS}}
\label{rel}
%Early detection of cancer significantly improves treatment outcomes and patient survival rates. However, nearly 50\% of cancers are still diagnosed at an advanced stage, highlighting the urgent need for more effective early detection strategies~\cite{crosby2022early}.Recent advances in biological understanding and technological progress have led to the development of innovative approaches to improve early diagnosis. The development of high-throughput sequencing technologies and ChIP-seq platforms has led to remarkable advancements in Single-Cell Sequencing (SCS) technology. This has significantly improved the diagnosis and treatment of various diseases, including cancer. However, the cost and time commitment are substantial, and the procedure is invasive since it involves the dissection of tumor tissue to isolate cells and conduct sequencing studies~\cite{han2022single}. In this context, the analysis and detection of Circulating Tumor Cells (CTCs) is crucial.The CTCs are rare cells shed from the primary tumor or metastatic sites into the bloodstream that have emerged as a valuable biomarker in oncology via liquid biopsy, providing a real-time and noninvasive means of monitoring disease progression and guiding therapeutic decisions~\cite{alix2021liquid}.Their extreme rarity and heterogeneity pose significant challenges for accurate identification and classification. However, detection of a specific marker may fail to capture the full spectrum of the heterogeneous CTC population, resulting in an underestimation of CTC counts in other clinical settings~\cite{wang2020label}. 
Several approaches have been proposed in the literature about CTC analysis. Many methods use images containing clusters of cells, and the aim is the detection and quantification of the cells. Guo et al. have proposed a CNN to automatically detect CTCs in peripheral blood using immunofluorescence in situ hybridization (imFISH) images~\cite{guo2022circulating}. However, this technique is based on counting the copy number of chromosome 8 using CEP8 immunofluorescence labeling rather than analyzing cell morphology.  BRIA (BReast cancer Imaging Algorithm), a fully automated ML-based pipeline designed to detect, segment, and classify metastatic breast CTC cells in multi-channel immunofluorescence images is proposed in~\cite{schwab2024fully}. Svensson et al.~\cite{svensson2014automated} presented a Naive Bayesian Classifier (NBC) to reliably and automatically detect and quantify CTCs. Cells are collected with a functionalized medical wire, stained for fluorescence microscopy, and the classification is performed by using RGB color histograms. In \cite{calvo2024multichannel}, the authors used a combination of computer vision and CNNs, that demonstrated strong performance in automating CTCs detection from multi-channel clustered fluorescence images, achieving high recall rates. 
%and reinforcing the potential of AI-driven methods to enhance early cancer detection and monitoring. 
Application on multi-channel images greatly improves the quality of the analyzed information, but the limited test dataset may not guarantee optimal generalization.

Given the scarcity and rarity of CTCs, Liang et al.~\cite{liang2024improving} used a novel data generation with Segment Anything Model (SAM) in combination with copy-paste to increase the number of images, which allows for improved generalization over existing models due to a new loss function. Despite the promising approach, the unreliability of synthetic data and computational cost remain challenges to consider.

A single-cell method is proposed in~\cite{zeune2020deep}, where an auto-encoder feed by fluorescent images of blood samples has been proved to have an accurate identification of CTCs. 

In \cite{miccio2020perspectives}, Miccio et al. demonstrated that label-free approaches eliminate the need for sample staining, reducing preparation time, potential alterations to cell properties, and dependence on specific dyes or markers.
In~\cite{liu2019faster} is presented a DL method that uses Whole Slide Images (WSI) for the automatic detection and enumeration of CTCs in microscopic blood images, reducing subjectivity and workload for cytologists. The high resolution of WSIs, however,  poses great challenges in their analysis and management \cite{brancati2021gigapixel}.

A tentative to merge information from fluorescence and BF images of single cells has been made in \cite{piansaddhayanon2023label}, where the authors have developed a proof-of-concept DL model that identifies cancer cells originating from cholangiocarcinoma in unlabeled microscopy images based on morphological differences, where each image contains 20–30 individual cells on average, emphasizing the importance of analysis on morphological features of cells that allow different cell types to be distinguished from each other. 

Wang et al.~\cite{wang2020negative} used immune microparticles to negatively mark white blood cells rather than CTCs, so that cancer cells can be directly distinguished in the BF channel of microscopy. In this way, all heterogeneous cancer cells and their phenotypic properties can be preserved for further cancer studies. However, applying such methodologies to clusters of cells could lead to the loss of crucial information. 
%For example, if we wanted to focus on studying the morphology of single cells, we would have to perform appropriate pre-processing of the images, which could lead to a significant loss of information. Therefore, the need arises to use methodologies that isolate and acquire single cells.

%The CTCs analysis is a challenging, intricate, and time-intensive manual task that requires experts to assess the images obtained from liquid biopsies individually. 
Akashi et al. \cite{akashi2023use} have demonstrated that, despite training their CNN-based detection system using KYSE520 cell lines with high EpCAM expression, the model has been still able to accurately identify KYSE30 cell lines, which are EpCAM-negative. This suggests that DL does not rely solely on traditional biomarkers, but can detect previously unknown or subtle features, such as cell morphology or minute differences in nuclear structure, to differentiate CTCs from blood cells.

The need to label the sample to obtain contrast agent effects greatly limits the generalization of methods based on fluorescence images, which can not be applied to datasets from different hospital structures. Moreover, analysis of single-cell images allows for a detailed examination of individual cell morphology, subcellular structures, and phenotypic variations that may be masked in clustered images.
It could be very useful to develop a method based on the analysis of BF images of single cells that overcomes the limitations of existing methods in the literature. 

\section{\textbf{METHODOLOGY}}
\label{met}
\subsection{\textbf{Isolation and characterization of CTCs}}
\begin{figure}
    \centering
    \includegraphics[width=1\linewidth]{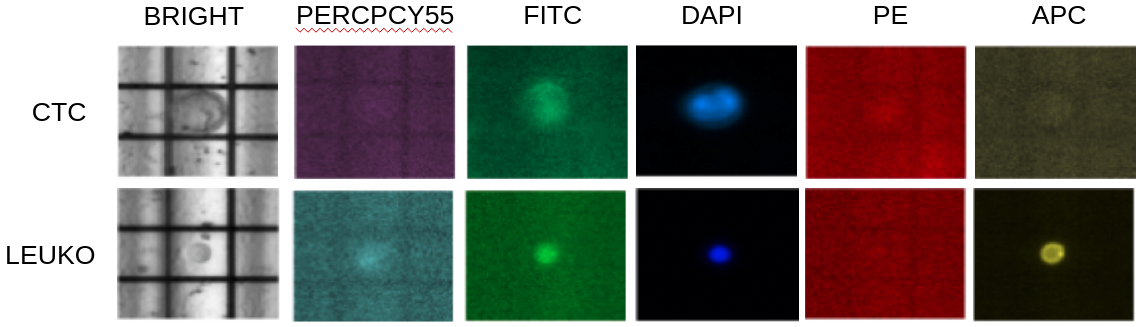}
    \caption{\textbf{Different types of images produced by DEParray.}}
    \label{fig:fig_out}
\end{figure}

Isolation and characterization of CTCs were performed according to a previously validated and described protocol [26]. Briefly, Peripheral Blood Mononuclear Cell (PBMC) were separated using SepMate™ PBMC Isolation Tubes from 10ml blood of two Healthy Voluntaries (HV), thirteen patients with Non-Small Cell Lung Cancer (NSCLC) and two with cholangiocarci- noma. For spike-in experiments, about 500 single cells from A549 and Calu3 NSCLC cell lines were spiked into PBMC of HV and processed as detailed below. PBMC from HV and patients were processed with Parsortix system to enrich for CTC. Recovered cells were stained with fluorescent antibodies against specific CTC markers and immune marker CD45, counter- stained with DAPI (4’,6-diamidino-2-phenylindole), used for nuclear staining which identifies cells with intact rounded nuclei. Fluorophore conjugated antibodies (detected in FITC/ PE/ PerCP-Cy5.5 channels) were used to analyze the expression of specific antigens potentially expressed by CTC and finally APC channel was used to identified CD45+ leukocytes. Representative images of spike-in cells were captured randomly for each fluorescent channel and BF. Putative CTCs, defined by cell size and negativity for CD45 expression, were isolated and subjected to DNA amplification, LowPass sequencing, and copy number alteration (CNA) analysis to confirm their tumor origin, as detailed in Vismara et al \cite{vismara2022single}.
Different types of images, stained and in grayscale, are produced by DE- PArray, see Fig. \ref{fig:fig_out}. The grayscale image is the BF channel, while the remain- ing images are the different channels stained with the specific markers. Not all channels were used to create the dataset for our experiments, but only images from BF and DAPI channels. We aim to streamline healthcare profes- sionals’ workflow by reducing the dependency on extensive sample staining with multiple markers.

\subsection{\textbf{Dataset Characteristics}}
\label{dataset}
The dataset comprises 529 images of tumor cell lines spiked into the blood of HV, resembling CTC, 52 CTC images from 13 patients with NSCLC, and 388 leukocyte images. The dataset was partitioned into three distinct subsets for the experiments: training, validation, and test sets. The training set was dedicated to learning the DL model, while the validation set served as an intermediate dataset to fine-tune hyperparameters during the model optimization process. The test set was reserved for evaluating the model’s performance on previously unseen data, mimicking real-world scenarios.
To enhance generalization during training and address the scarcity of real CTC images, spiked-in cell images were employed for both the training and validation phases, whereas actual CTC images were exclusively used for testing. For leukocytes, $15\%$ of the total images were allocated to the test set. In detail, 529 spiked-in cell images and 332 leukocyte images were utilized for training and validation, with $10\%$ of them randomly selected for validation. The final test phase included 52 images of CTCs and 56 images of leukocytes. (Table \ref{tab:my_label}).

\begin{table}
    \centering
     \caption{Number of dataset's images}
    \begin{tabular}{ccccc}
    \hline
        &\textbf{CTC}&\textbf{LEUKO} \\ 
        \hline
    Train&479&303 \\
     \hline
    Augmented Train&2395&1515 \\
     \hline
     Validation&50&29 \\
     \hline
    Test &52&56 \\
    \hline
   TOTAL&581&388 \\
     \hline
    \end{tabular}
   
    \label{tab:my_label}
\end{table}

\subsection{\textbf{Training Images}}
\label{preprocess}
Image augmentation techniques were utilized to expand the dataset and improve the model's ability to generalize. Specifically, three augmentation operations were implemented, incorporating random transformations like rotations and flipping, along with adjustments to brightness and color. The choice to apply these three types of augmentation was motivated by the goal of increasing data diversity. Additionally, images from the DAPI channel were included to further augment data, enhancing generalization and allowing the network to learn additional morphological CTCs' features. This multi-channel approach helps capture intrinsic characteristics of CTCs that may not be evident in BF images alone. Starting from a total of $782$ training images in the training set, a final number of $3910$ images was obtained after augmentation and the addition of DAPI channel images. In Table \ref{tab:my_label} the original and augmented training data are reported.

\subsection{\textbf{Proposed method}}
\begin{figure*}
    \centering
    \includegraphics[width=0.5\linewidth]{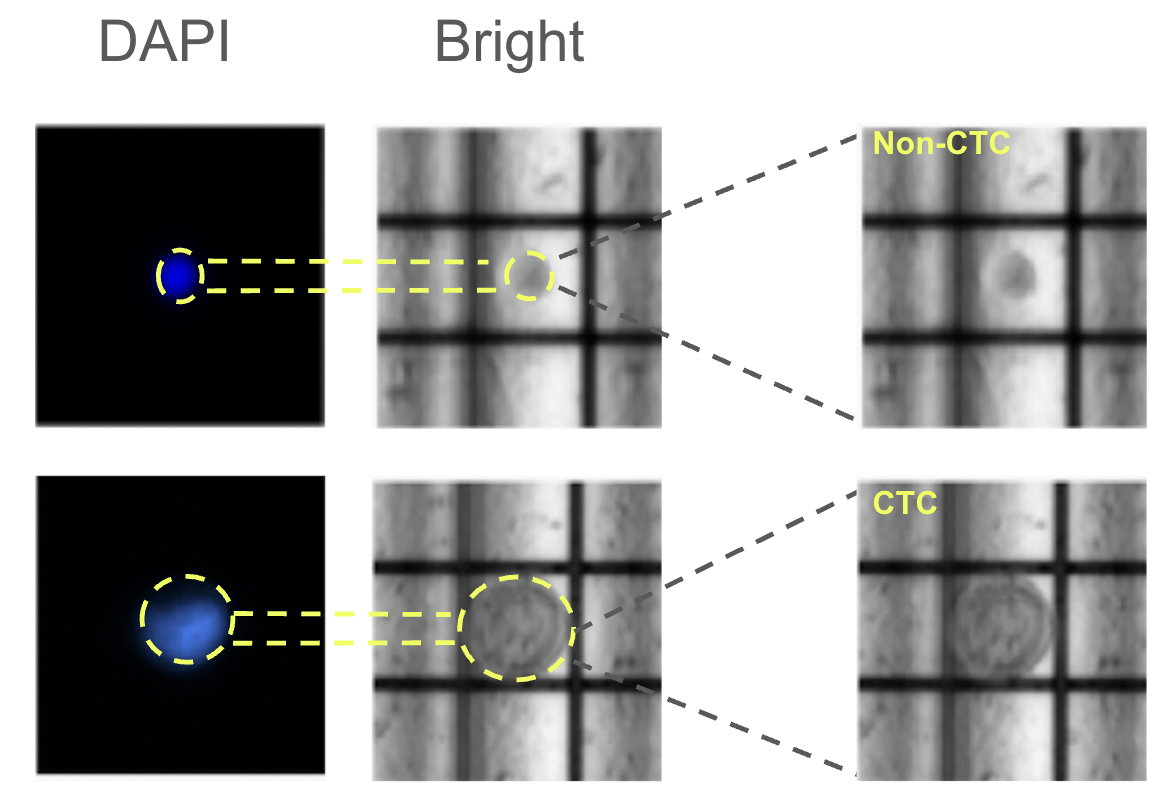}
    \caption{\textbf{Representative images of CTC and non-CTC cells in both DAPI and BF microscopy.}}
    \label{fig:image_microscopy}
\end{figure*}

The decision to rely on the BF channel is due to two main factors: i) it is a standard imaging output of DEPArray technology, and ii) the need to simplify the work of researcher operators by evaluating whether it is possible to avoid excessive staining of samples with different markers, which may also vary depending on both the type of tumor to be identified and the hospital structure. However, in BF images, background noise or artifacts may be more pronounced, particularly when the tissue or sample preparation is not ideal. This can interfere with the accurate interpretation of the image. Moreover, they often lack contrast, making it harder to distinguish subtle differences between cells, particularly when they have similar morphological characteristics. This can reduce the clarity and accuracy of cell identification. DAPI staining is relatively simple and inexpensive compared to other staining techniques. It does not require complex protocols, making it a convenient choice for routine use in various biological and medical imaging applications. The DAPI channel is highly specific and highly effective for visualizing cell nuclei and distinguishing nucleated cells from debris or anucleated cells, regardless of the tumor type. Representative images of CTC and non-CTC cells in both DAPI and BF microscopy are shown in Fig. \ref{fig:image_microscopy}.

Given these considerations, we decided to use BF in combination with DAPI channel images for our experiments. 
In particular, to exploit the DAPI property, we augment our training dataset by introducing DAPI images. In this way, we have a dual advantage: i) we increase the number of training images and ii) introduce new features (those from the DAPI images) that aid the learning process of the networks.  

On this dataset, we applied a DL approach, leveraging AI advancements to enhance classification and analysis of complex data. Specifically, we used CNNs, which are a type of neural network that is especially well-suited for image processing. These networks are structured to simulate the visual processing in the human brain, using convolutional layers to extract features from an image through specific filters. In other words, CNNs are capable of automatically learning patterns and relevant features from visual datasets without the need for manual feature extraction. In our study, we used several advanced CNNs architectures, including ResNet, \cite{he2016deep}, EfficientNet \cite{tan2019efficientnet}, and DenseNet \cite{huang2017densely}. These models were chosen due to their strong performance in handling complex image classification tasks, attributed to their efficiency in feature learning and ability to generalize across diverse datasets. ResNet, for instance, features a deep architecture with residual connections that help address the vanishing gradient problem. EfficientNet is specifically designed to optimize the trade-off between network depth, width, and input resolution, and finally, DenseNet employs densely connected layers to enhance information flow throughout the network.

Each architecture offers multiple variants tailored to different computational and performance needs. ResNet, for example, ranges from the lightweight ResNet18 with 18 layers to the more advanced ResNet152 with 152 layers. This diversity allows for flexibility in choosing models suited for various image recognition tasks.

A comparative analysis of these CNNs architectures was conducted, selecting specific versions that balance accuracy and computational efficiency, particularly for medical applications. Specifically, ResNet34 and ResNet50 were chosen from the ResNet family, while EfficientNetB4 and DenseNet121 were selected from the EfficientNet and DenseNet families, respectively. All networks were trained and tested on the dataset outlined in Section \ref{dataset}, incorporating the data augmentation strategies detailed in Section \ref{preprocess}.

\begin{figure*}
    \centering
    \includegraphics[width=1\linewidth]{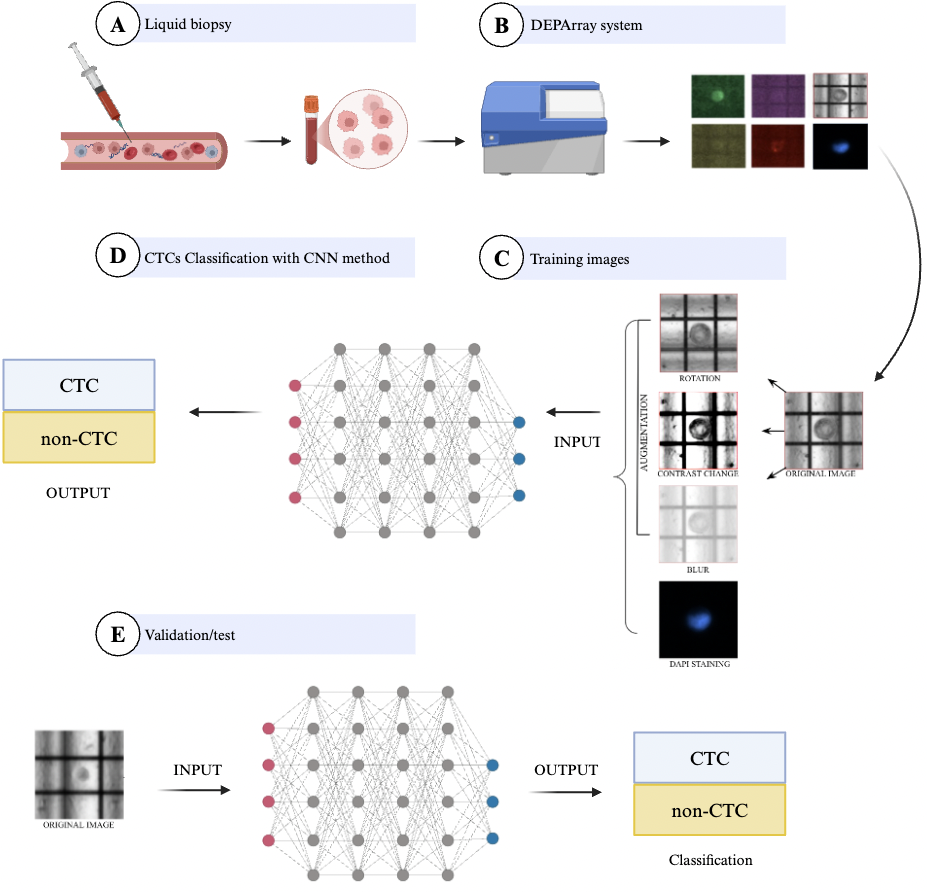}
    \caption{\textbf{Workflow of the proposed method.} \textbf{A.} Liquid biopsy is a blood test that detects cancer cells or tumor DNA, avoiding invasive procedures. \textbf{B.} The DEPArray™ system uses electric fields to isolate and select single cells, like CTCs. \textbf{C.} Training image pre-processing consist of BF DEPArray images whose variability is increased through the use of augmentation operations, and fluorescent-field images of DAPI-labeled cells. \textbf{D.} The output of the CNN is the classification of images into CTC and non-CTC. \textbf{E.} BF images are used in the validation/test phase to identify CTCs. }
    \label{fig:protocol}
\end{figure*}

The workflow of the proposed method is shown in Fig. \ref{fig:protocol}.

\section{\textbf{RESULTS AND DISCUSSION}}
\label{res}
\subsection{\textbf{Comparative results}}\label{ITH}
Several performance metrics were employed to evaluate the effectiveness of classification models, including Accuracy, F1-score, Precision, and Recall. These metrics rely on four key components: true positives (TP), which represent correctly identified positive cases; false positives (FP), where negative instances are mistakenly classified as positive; true negatives (TN), referring to correctly identified negative cases; and false negatives (FN), where positive instances are incorrectly classified as negative.
Accuracy measures the proportion of correct predictions out of the total, offering a broad assessment of model performance. However, in cases of class imbalance, accuracy alone can be misleading. The F1-score, calculated as the harmonic mean of Precision and Recall, provides a balanced evaluation by considering both false positives and false negatives. This makes it particularly useful when dealing with imbalanced datasets where neither Precision nor Recall should outweigh the other. Precision focuses on how many of the predicted positive cases are actually correct, which is critical in situations where false positives carry significant consequences, such as in medical screenings or fraud detection. Conversely, Recall measures how well the model identifies actual positive cases, making it especially important in scenarios where missing a positive instance, such as in disease diagnosis, could have serious repercussions.

We evaluated these metrics on several CNNs, including DenseNet121, EfficientNetB4, ResNet50, and ResNet34, all pre-trained on the ImageNet dataset. Considering the imbalance of the training dataset, the F1-score was chosen as the metric to evaluate the proposed method. So, for each architecture, the best model with the highest F1-score was chosen during the validation phase, and then this model was evaluated on a test set. 
The images were all resized to 148×148, which corresponds to the size of the smallest image in the dataset. Cross entropy loss was considered for the backpropagation and an AdamW optimizer~\cite{loshchilov2017decoupled} was adopted with an initial learning rate equal to $10^{-7}$.
For each experiment, the mean and standard deviation of the selected evaluation metrics on the test set using five random weight initializations were computed. 
%The F1 scores achieved by these models were as follows: DenseNet121 ($0.420$), EfficientNetB4 ($0.519$), ResNet50 ($0.798$), and ResNet34 ($0.451$). 
The average performance metrics are summarized in Table \ref{comparison}.
ResNet50 achieved the highest F1 score ($0.798$), indicating that this model significantly outperformed the others in terms of the balance between precision and recall. This suggests that ResNet50 was particularly effective at learning the relevant features for the task.
With an F1 score of $0.578$, EfficientNetB4 performed better than DenseNet121 and ResNet34, but still lagged behind ResNet50. EfficientNet models are designed to be more parameter-efficient, but in this case, it seems that EfficientNetB4 did not generalize as well as ResNet50. Despite belonging to the same ResNet family, ResNet34 achieved an F1 score of $0.407$, which is much lower than ResNet50. This suggests that the deeper architecture of ResNet50 allowed it to learn more complex features, leading to significantly better performance. The lowest F1 score ($0.378$) was observed with DenseNet121. While DenseNet architectures are known for efficient feature reuse, this result indicates that DenseNet121 struggled with the dataset, possibly due to overfitting, or difficulties in extracting discriminative features.

Unfortunately, a direct comparison with existing studies in the literature is not feasible, as most approaches focus on different types of images, such as fluorescence-based imaging or cell clusters. Consequently, the architectures proposed in those studies are not directly applicable to our framework, which relies solely on BF images and single-cell analysis. On the other hand, it is not possible to validate our architecture on their datasets, as the corresponding images are not publicly available online.
\begin{table*}[]
\centering
\caption{Results of the different CNN architectures. In bold the best results for each measure.}
\resizebox{\textwidth}{!}{
\begin{tabular}{ccccc}
\hline
                                         & \textbf{Accuracy} & \textbf{Precision} & \textbf{Recall} & \textbf{F1-score} \\ 
                                         \cline{1-5}
\multicolumn{1}{l}{\textbf{EfficientnetB4}}  & $0.579  \pm 0.027$     & $0.580  \pm 0.024$     & $0.580  \pm 0.024$       & $0.578  \pm 0.027$          \\
\hline
\multicolumn{1}{l}{\textbf{Densenet121}}   & $0.535 \pm 0.005$        & $0.760 \pm 0.000$    & $0.518 \pm 0.005$            & $0.378 \pm 0.011$    \\
\hline
\multicolumn{1}{l}{\textbf{Resnet34}}     & $0.549  \pm 0.014$     & $0.765  \pm 0.005$          & $0.533  \pm 0.015$    & $0.407  \pm 0.029$        \\
\hline
\multicolumn{1}{l}{\textbf{Resnet50}}         & $\textbf{0.798}  \pm \textbf{0.003}$    & $\textbf{0.828}  \pm \textbf{0.011}$               & $\textbf{0.804}  \pm \textbf{0.005}$       & $\textbf{0.798}  \pm \textbf{0.005}$ \\ \hline    
\end{tabular}}
\label{comparison}
\end{table*}

\subsection{\textbf{Ablation study}}\label{ITH}
To assess the importance of the introduction of DAPI images in the augmentation process, an ablation study was performed. In Table \ref{tab_aug_dapi_bf}, the results of the following experiments are reported:
\begin{itemize}
    \item AUG1: only one augmentation operation;
    \item AUG2: two augmentation operations;
    \item BF w/o DAPI: three augmentation operations, but without DAPI images;
    \item BF w/ DAPI no AUG: augmentation operation only with DAPI images;
    \item BF w/ DAPI: our proposed approach, with three augmentation operations and a further augmentation with DAPI images.
\end{itemize}
Observing the results, it is possible to note that a key element in the enhanced performance of the ResNet50 model was the implementation of carefully designed data augmentation strategies. These techniques substantially increased the diversity of the training set, addressing the challenges associated with the limited initial dataset. By introducing a broader range of variations, the model was better equipped to learn robust and generalizable features, ultimately improving its classification accuracy and reducing the risk of overfitting. In particular, the results improved as the number of augmentation operations introduced increased (AUG1, AUG2). Indeed, the application of three specific augmentation functions improved the robustness of the model (BF w/o DAPI). Furthermore, to understand the importance of augmentation operations during training, we attempted to train the model without them, inserting only DAPI images to increase the variability of the data (BF w/ DAPI no AUG). These findings indicate that a substantial number and variability of data are essential for optimal model performance. Finally, integrating an auxiliary dataset containing DAPI fluorescence channel images in the training set further enhanced performance across all evaluated metrics (BF w/ DAPI). 

To better confirm the importance of BF image analysis and that DAPI images were used only for additional support, we also conducted experiments by reversing the roles of the two types of images:
\begin{itemize}
    \item DAPI w/o BF: DAPI images were used as main images for the analysis, and three augmentation operations were performed on them for the training phase. In the validation and test phase, only DAPI images were used for the final classification;
    \item DAPI w/ BF: for the training phase, DAPI images, three augmentation operations on them, and BF images were used. As before, only DAPI images were used for the final classification in the validation and test phase.
\end{itemize}
The results in this case were rather discouraging in both cases: for DAPI w/o BF, the F1-score decreased from $0.798$ to $0.510$, while for DAPI w/ BF the F1-score even decreased to $0.474$.
These results show that the utilization of images derived from the BF during the testing phase is of paramount importance. These images enable the model to analyze the morphological shape of the cell, facilitating its classification into the appropriate category. On the other hand, the utilization of images in the fluorescence field of DAPI was demonstrated to be pivotal during the model's training phase, as it enables the network to extract the crucial information necessary for the subsequent validation step.      

\begin{table*}[h]
\centering
\caption{Results of Ablation study}
\resizebox{\textwidth}{!}{
\begin{tabular}{ccccc}
\hline
                   & \multicolumn{1}{c}{\textbf{Accuracy}} & \multicolumn{1}{c}{\textbf{Precision}} & \multicolumn{1}{c}{\textbf{Recall}} & \multicolumn{1}{c}{\textbf{F1-Score}} 
                   \\ \hline
\textbf{AUG1}  & $0.722 \pm 0.009$ & $0.793 \pm 0.015$ & $0.730 \pm 0.010$ & $0.709 \pm 0.010$                                                   \\ \hline
\textbf{AUG2}  & $0.759 \pm 0.009$ & $0.793 \pm 0.012$ & $0.767 \pm 0.006$ & $0.757 \pm 0.010$    
\\ \hline

\textbf{BF w/o DAPI}   & $0.780 \pm 0.012$ & $0.806 \pm 0.017$ & $0.782 \pm 0.013$ & $0.777 \pm 0.012$                                                    \\ \hline

\textbf{BF w/ DAPI no AUG}   & $0.663 \pm 0.017$ & $0.690 \pm 0.026$ & $0.660 \pm 0.010$ & $0.641 \pm 0.011$                                                   \\ \hline
\textbf{BF w/ DAPI}  & $\textbf{0.798} \pm \textbf{0.003}$ & $\textbf{0.828} \pm \textbf{0.011}$ & $\textbf{0.804} \pm \textbf{0.005}$ & $\textbf{0.798} \pm \textbf{0.005}$\\ \hline

\textbf{DAPI w/o BF}  & $0.537 \pm 0.009$ & $0.560 \pm 0.010$ & $0.550 \pm 0.010$ & $0.510 \pm 0.010$                                                   \\ \hline
\textbf{DAPI w/ BF}  & $0.497 \pm 0.019$ & $0.507 \pm 0.021$ & $0.507 \pm 0.021$ & $0.474 \pm 0.027$                                                   \\ \hline

\end{tabular}}
\label{tab_aug_dapi_bf}
\end{table*}
\subsection{\textbf{Statistical analysis}}\label{STR}
\begin{figure*}
    \centering
    \includegraphics[width=1\linewidth]{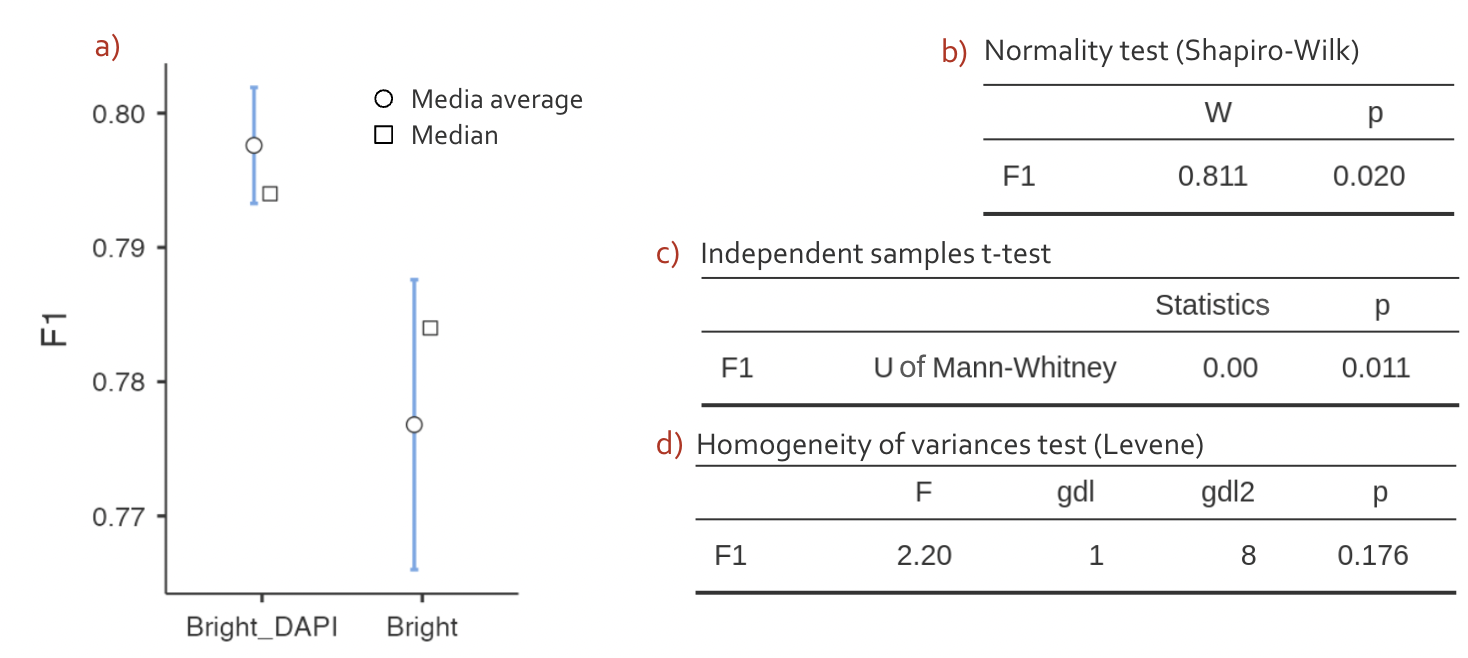}
    \caption{\textbf{Statistical results of Mann-Whitney test} a) Statistical results with $p-value<0.05$ demonstrate a statistically significant difference between the two groups considered (BF and BF with DAPI). b) The data do not follow a normal distribution. c) Mann-Whitney test. d) Homogeneity test.}
    \label{fig:statistics}
\end{figure*}

A statistical analysis of the data has been performed. We used Jamovi software to conduct Student's t-test. The data were analyzed by checking the homogeneity of variance using Levene's test ($p-value=0,176$), and the the normality of the distributions using the Shapiro-Wilk test, which showed that the data did not follow a normal distribution since the p-value is lower than $0.05$ ($p-value=0,020$). So, considering this not normal distribution and the sample size was relatively small, the Mann-Whitney U test, a non-parametric test that does not assume normality, was employed. In particular, to evaluate the impact of including images in the DAPI channel on the model's performance during the training set, the Mann-Whitney U test was used to compare the F1 score obtained under two conditions: (1) the use of the dataset containing BF images with three augmentation operations, and (2) the use of the combined dataset containing both BF with three augmentation operations and DAPI channel images (BF w/o DAPI vs BF w/ DAPI). The F1-score was chosen as the evaluation metric since it balances precision and recall, providing an overall measure of the model's performance. The Mann-Whitney U test was performed to determine if the difference in F1-score between the two groups was statistically significant, with a significance level set at $0.05$ ($p-value=0,011$) (see Fig. \ref{fig:statistics}).
This result indicates that the inclusion of the dataset containing DAPI channel images as an augmentation operation is crucial for increasing data variability, ultimately enhancing the model’s generalization capabilities. The observed statistical significance strongly supports the rejection of the null hypothesis, confirming that this improvement is not due to random variability but rather to the contribution of fluorescence-derived features. Moreover, training solely on BF images, even with augmentation, appears insufficient to achieve comparable performance, highlighting the necessity of leveraging fluorescence-enhanced training strategies. These findings reinforce the practical importance of incorporating additional feature-rich data sources during training, ensuring a more robust and reliable classification pipeline for CTC identification. 
%In a broader context, this improvement could translate into more accurate liquid biopsy-based cancer detection, contributing to earlier diagnosis and more effective patient management.

\section{\textbf{CONCLUSION}}
\label{con}
CTCs play a crucial role as biomarkers in liquid biopsy, providing a minimally invasive tool to monitor cancer progression and guide therapeutic strategies. Despite their clinical potential, the identification of CTCs remains challenging due to their extreme rarity and heterogeneity. To address these issues, an unbiased workflow combining Parsortix® technology, which isolates CTCs based on size and deformability, with DEPArray™ technology, which enables the precise visualization and selection of single cells, has been developed. Traditionally, the analysis of cell images obtained through DEPArray is performed manually by experienced reaserchers, a process that is both time-intensive and prone to variability.
Automating the identification of CTCs within clinical workflows could streamline the detection of metastases, enhance the accuracy of therapeutic decisions, and ultimately improve patient outcomes. In this study, we have presented a DL-based classification system designed to differentiate CTCs from leukocytes in liquid biopsy samples. The system leverages the ResNet architecture, a CNN widely recognized for its robustness in medical image analysis.
Among the tested architectures, applied to images acquired with DEPArray technology, ResNet50 demonstrated superior performance, achieving an F1-score of $0.798$, which was statistically significant ($p-value < 0.05$) compared to ResNet34 ($0.407$), EfficientNetB4 ($0.578$), and DenseNet121 ($0.378$). This result underscores the suitability of ResNet50 for addressing the inherent complexity of BF images of CTCs.
A pivotal element in achieving these promising results was the implementation of tailored data augmentation techniques, which increased the variability of the training data and compensated for the limited size of the initial dataset. Specifically, three augmentation functions were employed to improve the robustness of the model, and further augmentation was given by the inclusion of an additional dataset images of the DAPI fluorescence channel led to further improvements in all performance metrics. %Moreover, the inclusion of an additional dataset consisting of DAPI fluorescence channel images led to further improvements across all performance metrics.
These findings highlight the potential of combining data augmentation strategies with multimodal imaging to overcome dataset limitations, advancing the application of DL in CTCs identification. Future work will focus on segmenting CTCs images to extract key morphological features that distinguish them from leukocytes.

\section*{Acknowledgment}

This work was funded by European Union – Next Generation EU, Missione 4, Componente 1, “Fondo PRIN 2022” – HIMALAYA – ID 20228MHWPZ  - CUP B53D23002380006, whose support was instrumental in carrying out this study.

\end{document}